\documentclass[%
 onecolumn,
 notitlepage,
showpacs,%preprintnumbers,
 amsmath,amssymb,
 aps,
pra,
]{revtex4-1}

\usepackage{graphicx}% Include figure files
\usepackage{dcolumn}% Align table columns on decimal point
\usepackage{bm}% bold math
\usepackage{hyperref}% add hypertext capabilities
\usepackage{braket} % Allows for Dirac notation

\newcommand{\threej}[6]{ \begin{pmatrix}
  #1 & #2 & #3 \\
  #4 & #5 & #6 
 \end{pmatrix}}

\newcommand{\sixj}[6]{ \begin{Bmatrix}
  #1 & #2 & #3 \\
  #4 & #5 & #6 
 \end{Bmatrix}}

\newcommand{\ninej}[9]{ \begin{Bmatrix}
  #1 & #2 & #3 \\
  #4 & #5 & #6 \\
  #7 & #8 & #9
 \end{Bmatrix}}

\begin{document}

%\preprint{APS/123-QED}

\title{Energy spectra of two interacting fermions with spin-orbit coupling in a harmonic trap}% Force line breaks with \\

\author{Cory D. Schillaci}
\email{schillaci@berkeley.edu}
\affiliation{%
Department of Physics, University of California, Berkeley, California 94720, USA
}%

\author{Thomas C. Luu}%
\email{
t.luu@fz-juelich.de
}
\affiliation{
Institute for Advanced Simulation, \\
Institut f\"{u}r Kernphysik and J\"{u}lich Center for Hadron Physics, Forschungszentrum J\"{u}lich, D-52425 J\"{u}lich, Germany
}

\date{December 17, 2014}% It is always \today, today,

\begin{abstract}
We explore the two-body spectra of spin-$1/2$ fermions in isotropic harmonic traps with external spin-orbit potentials and short range two-body interactions. Using a truncated basis of total angular momentum eigenstates, nonperturbative results are presented for experimentally realistic forms of the spin-orbit coupling: a pure Rashba coupling, Rashba and Dresselhaus couplings in equal parts, and a Weyl-type coupling. The technique is easily adapted to bosonic systems and other forms of spin-orbit coupling.
\end{abstract}

\maketitle

%\tableofcontents

\section{\label{sec:level1}Introduction}

Cold atomic gases with spin-orbit coupling (SOC) have recently been an area of intense interest because of the potential to simulate interesting physical systems with precisely tunable interactions \cite{nature11841}. In condensed matter physics, spin-orbit couplings are essential for many exotic systems such as topological insulators \cite{das2013engineering,PhysRevLett.105.255302}, the quantum spin Hall effect \cite{nature12185}, and spintronics \cite{RevModPhys.76.323}. The experimental setup which induces spin-orbit coupling is intimately related to simulation of synthetic gauge fields \cite{RevModPhys.83.1523,hamner2014dicke,Lin:2009zzb,Bermudez:2011db}. Because these couplings are parity violating, they potentially play similar roles within nuclear systems that undergo parity-violating transitions due to the nuclear weak force.  Atomic gases provide an excellent testing ground both to explore universal behavior of these real life systems and to create new types of spin-orbit coupling which are not yet known to exist (or have no solid-state analog) in other materials but are interesting in their own right. Further, these experiments can be performed in an environment with few or no defects and impurities.

Spin-orbit coupling was first realized in a Bose condensate of $^{87}$Rb \cite{nature09887} and extended shortly after to Fermi gases of $^{40}$K \cite{PhysRevLett.109.095301} and $^6$Li \cite{PhysRevLett.109.095302}. These spin-orbit interactions are `synthetic' in the sense that a subset of the hyperfine states stand in as virtual spin states. A particularly interesting consequence of this is the possibility of studying systems with synthetic spin-$1/2$ spin-orbit interactions but bosonic statistics \cite{PhysRevA.68.063612,nature09887}. From another point of view, the couplings are equivalent to applying external electromagnetic forces via synthetic gauge couplings on the physically uncharged particles in the gas \cite{Lin:2011,PhysRevLett.107.255301}. It has also been conjectured that these systems could be used to physically simulate lattice gauge theories \cite{Bermudez:2010da,Mazza:2011kf}.  Spin-orbit couplings in solid-state systems arise in two-dimensional (2D) systems (Rashba and Dresselhaus types, described in Sec.~\ref{sec:Hamiltonian}), but recently an experimental setup has been proposed that can simulate the Weyl-type SOC which is fundamentally three dimensional \cite{PhysRevLett.108.235301}.

Spin-orbit couplings are also of interest from the perspective of few-body physics where they arise in a variety of fields, e.g., the weak nuclear interactions governing proton-proton scattering \cite{Haxton:2013aca,deVries:2014vqa}. Because the spin-orbit coupling is long range, it can significantly modify both the threshold scattering behavior and the spectrum of two-body systems \cite{PhysRevA.86.042707}. For low-energy scattering, Duan \textit{et al.} \cite{PhysRevA.87.052708} showed analytically that parity-violating SOC leads to the the spontaneous emergence of handedness in outgoing states, a finding later confirmed in \cite{PhysRevA.91.022706}. Even in the presence of a repulsive two-body interaction, an arbitrarily weak SOC has been shown to bind dimers \cite{PhysRevB.83.094515}. For three-particle systems, a new type of universality is conjectured to occur for bound trimers with negative scattering length \cite{PhysRevLett.112.013201}. 

Few-atom systems undergoing SOC within trapping potentials have also been explored. For example, the spectrum of particles within a trap with an external SOC of the Weyl type (but no relative interaction) has been theoretically determined \cite{anderson2013}. The Rashba SOC with two-particle systems interacting via short-ranged interactions was investigated perturbatively in \cite{PhysRevA.89.033606}, where it was shown that the leading order corrections due to the SOC and short-range interaction are independent when the scattering length is equal for all channels.  In one dimension, the spectrum for this type of system has been calculated when the SOC consists of equal parts Rashba and Dresselhaus interactions \cite{guan2014energy}. Information learned from trapped systems augments that from scattering experiments while also being relevant to interesting phenomena in trapped many-body systems with SOC such as solitons \cite{DarkSolitons,PhysRevA.87.013614} or novel phase diagrams \cite{PhysRevLett.107.270401}.

In all these calculations, the emergent spectrum is rich and complex, offering new insights into few-body behavior.  Our objective is to provide some additional insight into two-body physics of Fermi gases with spin-orbit interactions in the presence of both three-dimensional trapping potentials and short-ranged two-body interactions, which are necessarily present in dilute cold-atom experiments. Our approach is to numerically diagonalize the Hamiltonian within a suitably truncated basis, and is thus nonperturbative in nature. Eigenstates of the interacting Hamiltonian without SOC are used for the basis. Section~\ref{sec:Hamiltonian} introduces the specific forms of spin-orbit coupling and two-body interactions which we consider. The general method is detailed in Sec.~\ref{sec:Weyl} for the simplest SOC.  In the remaining Secs.~\ref{sec:Rashba}-\ref{sec:R=D} we study the spectra of additional spin-orbit couplings in order of increasing computational complexity.

\section{\label{sec:Hamiltonian}Hamiltonian for Spin-orbit Couplings with Contact Interactions}

In this paper we simply refer to our systems by their `spin' degrees of freedom and use the standard notation for spin quantum numbers. We consider three different types of spin-orbit coupling. The form of spin-orbit coupling realized in experiments is a linear combination of the Rashba \cite{0022-3719-17-33-015} and linear Dresselhaus \cite{PhysRev.100.580} types,
\begin{align}
V_{R}&\equiv\alpha_R (\sigma_x k_y-\sigma_y k_x) \label{eq:Rashba},\\
V_{D}&\equiv\alpha_D (\sigma_x k_y+\sigma_y k_x) \label{eq:Dresselhaus},
\end{align} 
which were originally recognized in two-dimensional solid-state systems. In a 2D system, these form a complete basis for spin-orbit couplings linear in momentum. Note that some references use the alternate definitions $V_R\propto  (\sigma_x k_x+\sigma_y k_y) $ and $V_D\propto  (\sigma_x k_x-\sigma_y k_y) $ which are equivalent up to a pseudospin rotation.  For solids, these parity-violating interactions are allowed only in the absence of inversion symmetries. Rashba-type SOC typically arises in the presence of applied electric fields or in 2D subspaces such as the surfaces of materials where the boundary breaks the symmetry. Dresselhaus couplings were first studied in the context of bulk inversion asymmetry, when the internal structure leads to gradients in the microscopic electric field. 

To date, experiments have produced only SOC potentials in which the Rashba and Dresselhaus terms appear with equal strength (also known as the ``persistent spin-helix symmetry point'' \cite{PhysRevLett.97.236601}), 
\begin{equation}
\label{eq:R=D}
V_{R=D}\equiv\alpha_{R=D}\sigma_x k_y.
\end{equation} 
After a pseudospin rotation, this potential can be seen as a unidirectional coupling of the pseudospin and momentum along a single axis. A proposal for tuning the ratio $\alpha_R/\alpha_D$ has been given in \cite{PhysRevA.84.025602}.  An experimental setup which gives the simple three-dimensional Weyl coupling,
\begin{equation}\label{eq:Weyl}
V_{W}\equiv\alpha_W \vec{k}\cdot\vec{\sigma},
\end{equation}
has also been proposed in \cite{PhysRevLett.108.235301} and \cite{PhysRevLett.111.125301}. 

In the following sections we calculate the spectra of two particles with a short-range two-body interaction, an isotropic harmonic trapping potential and spin-orbit coupling. The single particle Hamiltonian is 
\begin{equation}\label{eq:shortRangeInteraction}
H_1=\frac{\hbar^2 k^2}{2m}+\frac{1}{2}m\omega^2 r^2 + V_{\text{SO}}.
\end{equation}
For the spin-orbit term $V_{\text{SO}}$, we consider equal Rashba and Dresselhaus~\eqref{eq:R=D}, pure Rashba~\eqref{eq:Rashba}, and Weyl~\eqref{eq:Weyl} spin-orbit couplings  because these are generally considered to be experimentally feasible.

We assume that the range of interaction between particles is small compared to the size of the oscillator well.  The relative interaction between the particles can then be approximated as a regulated $s$-wave contact interaction, which in momentum space (as a function of relative momentum) is given by
\begin{equation}
\frac{4\pi \hbar^2}{m}a(\Lambda)\ .
\end{equation}
Here the argument $\Lambda$ refers to some cutoff scale and $a(\Lambda)$ is some function of the cutoff and physical scattering length $a_{\text{phys}}$.  The exact form of this function depends on the type of regulator used and is not relevant for this work; the only constraint is that $a(\Lambda)$ reproduce the physical scattering length given by the scattering $T$ matrix at threshold, $T(E=0)=4\pi\hbar^2 a_{\text{phys}}/m$ \cite{taylor2000}. In the limit $\Lambda\rightarrow \infty$ the spectrum of two particles in an oscillator well (without external spin-orbit interaction) was solved by Busch \textit{et al.} \cite{Busch} using the method of pseudopotentials.  In Ref.  \cite{Luu:2006xv} the solution for general $\Lambda$ was given using a Gaussian regulator, which in the limit $\Lambda\rightarrow\infty$ recovered the Busch \textit{et al.} solution.  For our work below we use the eigenstates and eigenvalues of this two-particle system given in Ref. \cite{Busch}.

\section{\label{sec:Weyl}Weyl coupling}
We tackle the Weyl form first because of its mathematical and numerical simplicity. In the absence of the two-body interaction, this problem was treated in Ref. \cite{anderson2013}. Our approach is to determine the matrix elements of the SOC in an appropriate basis. The eigenvalue is then solved numerically at the desired precision by choosing an appropriately large truncated basis of harmonic oscillator (HO) eigenstates.

As usual, the two-body problem is best approached in the dimensionless Jacobi coordinates
\begin{equation}
R=\frac{r_1+r_2}{\sqrt{2}b}, \qquad r=\frac{r_1-r_2}{\sqrt{2}b}
\end{equation}
and the corresponding conjugate momenta $q,Q$ representing the relative and total momenta. For an isotropic harmonic oscillator, distances can be expressed in terms of the ground-state length scale $b=\sqrt{\hbar/m\omega}$ and energies will be similarly measured in units of $E_0=\hbar\omega$. We also define the spin operators
\begin{equation}
\vec{\sigma}\equiv\vec{\sigma}_1-\vec{\sigma}_2, \qquad \vec{\Sigma}\equiv\vec{\sigma}_1+\vec{\sigma}_2.
\end{equation}

With these definitions, the two-body Hamiltonian can be nondimensionalized and separated into relative and center-of-mass (c.m.) parts,
\begin{equation}\label{eq:WeylHamiltonian}
\frac{1}{\hbar\omega}H=\left(h_{0,\text{rel}}+\frac{\tilde{\alpha}_W}{\sqrt{2}} \vec{q}\cdot\vec{\sigma} + \sqrt{2}\pi \tilde{a}(\Lambda) \delta^{(3)}(r)\right)+\left(h_{0,\text{c.m.}}+\frac{\tilde{\alpha}_W}{\sqrt{2}} \vec{Q}\cdot\vec{\Sigma} \right),
\end{equation}
where $h_{0,\text{rel}}=r^2/2$ and $h_{0,\text{c.m.}}=R^2/2$. Notably, the spin-orbit coupling appears in both terms.  The tilde over the coupling constants indicates that they are dimensionless, related to the original coupling constants by dividing out the oscillator length (e.g., $\tilde{\alpha}=\alpha/b$). Throughout the remainder of this paper we will refer to dimensionless eigenvalues of $H/\hbar\omega$ as the energies of the system.

%We point out that the relative-coordinate spin-orbit term in Eq.~\eqref{eq:WeylHamiltonian} is exactly of the form that appears in weak-interaction parity-violating proton-proton scattering \cite{Haxton:2013aca,deVries:2014vqa}.  Aside from Coulomb contributions, the $^1S_0$ channel of proton-proton scattering has a scattering length that is an order of magnitude larger than its effective range.  The parity-conserving part of the nuclear potential could therefore be represented by the contact interaction in Eq.~\eqref{eq:WeylHamiltonian}.  The CM spin-orbit term (i.e. the last term in Eq.~\eqref{eq:WeylHamiltonian}) spoils this analogy. However, we will show later that the effect of this term on the ground state is negligible. 

\begin{figure}
\includegraphics{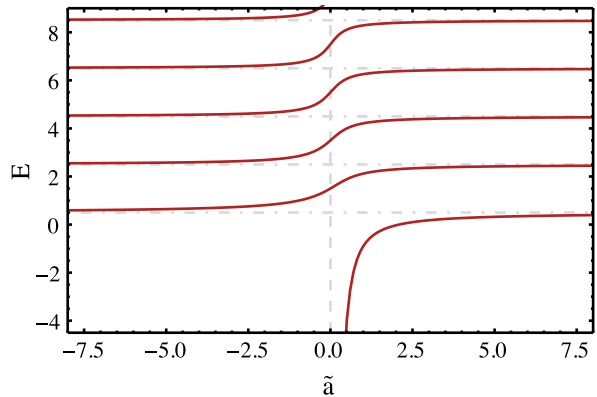}
\caption{\label{fig:BuschSpectrum}(Color online) Spectrum of the two-body contact interaction Hamiltonian as a function of $\tilde a$. The horizontal lines indicate the dimensionless energy eigenvalues in the unitary limit $|\tilde{a}|\rightarrow\infty$.} 
\end{figure}

Eigenstates of two particles with a short-range interaction in a harmonic oscillator trapping potential form a convenient basis for these calculations. These basis functions were first derived in \cite{Busch} for the isotropic case considered here, and the more general case of an anisotropic trap has been explored in \cite{PhysRevA.74.022712}. The dependence of the energy spectrum on the scattering length $a$ is shown in Fig.~\ref{fig:BuschSpectrum} for reference. Qualitatively, the effect of the short-range interaction is to shift the harmonic oscillator energies by $\pm \hbar\omega$ as the scattering length goes to $\pm \infty$. For positive scattering length, there is also an additional negative-energy dimer state.

We choose the particular coupling scheme of angular momentum eigenstates,
\begin{equation}\label{eq:basisStates}
\ket{n(ls)j;NL;(jL)J},
\end{equation}
which simplify the matrix elements for the relative-coordinate operators. Here $n$ and $l$ refer to the principal and orbital angular-momentum quantum numbers of the two-particle system in the relative coordinates. $N$ and $L$ refer to the analogous numbers in the center-of-mass frame. The total spin of the two spin-$1/2$ particles is denoted by $s = s_1 + s_2$ and may be either 0 or 1. First $s$ and $l$ to make angular momentum $j$, which is then recoupled with the c.m. angular momentum $L$ to make the state's total angular momentum $J$. Because all terms in the Hamiltonian~\eqref{eq:WeylHamiltonian} are scalars, the interaction is independent of $J_z$ and so we omit this quantum number for clarity. Due to Pauli exclusion, $l + s$ must be even to enforce antisymmetry under exchange of the particles.

For $l\neq0$ the states~\eqref{eq:basisStates} are identical to the well known harmonic oscillator, with $n$ and $l$ ($N$ and $L$) indicating the relative (center-of-mass) HO quantum numbers. We use the convention that $n,N=0,1,2,\dots$, and therefore $E=2n+l+2N+L+3$. The short range interaction~\eqref{eq:shortRangeInteraction} modifies the $l=0$ states and their spectrum. The principal relative quantum number $n$ for these states is obtained by solving the transcendental equation
\begin{equation}\label{eq:eigenvalueEqn}
\sqrt{2}\frac{\Gamma(-n)}{\Gamma(-n-1/2)}=\frac{1}{a}
\end{equation}
and is no longer integer valued. For the relative-coordinate part of the $l=0$ wave function,
\begin{align}
\phi(r)&=\frac{1}{2\pi^{3/2}}A(n)\Gamma(-n)U(-n,3/2,r^2)e^{-r^2/2}, \label{eq:BuschWF}\\
A(n)&=\left(\frac{\Gamma(-n)[\psi_0(-n)-\psi_0(-n-1/2)]}{8 \pi^2 \Gamma(-n-1/2)}\right)^{-1/2},
\end{align}
where $U(a,b,x)$ is Kummer's confluent hypergeometric function and $\psi_0(x)=\Gamma'(x)/\Gamma(x)$ is the digamma function. A derivation of the normalization factor $A(n)$ is given in the Appendix.

Standard angular momentum algebra can be used to determine the matrix elements of the two spin-orbit coupling terms; we follow the conventions of \cite{Edmonds}. For Weyl SOC of two spin-$1/2$ fermions, the matrix elements of the coupling in the relative momentum are
\begin{equation}\label{eq:WeylRel}\begin{split}
\bra{n'(l's')j';N'L';(j'L')J'}\vec{q}&\cdot\vec{\sigma} \ket{n(ls)j;NL;(jL)J}  \\
=&\delta_{N,N'}\delta_{L,L'}\delta_{j,j'}\delta_{J,J'}(-1)^{l+s'+j}\frac{3}{\sqrt{2}}\sixj{j}{s'}{l'}{1}{l}{s} (s'-s)\braket{n'l' || q || n l}.
\end{split}
\end{equation}
To preserve anti-symmetry of the two-particle system, the relative momentum term in the Weyl SOC must couple states with relative angular momentum $l$ to $l\pm 1$, leaving $l+s$ even but changing the parity.

For basis states with both $l,l'\neq0$, reduced matrix elements of the momentum operator are calculated between pure harmonic oscillator states,
% Could do away with some constants in favor of the 3-j coefficient
\begin{align}
\braket{n'l' || q || n l}=&(-1)^{l'}(-1)^{\frac{l+l'+1}{2}}\sqrt{\frac{2(2l+1)(2l'+1)}{(l+l'+1)}}\braket{n'l'0| (-i \nabla_0) | n l 0} \\
\begin{split} =& i(-1)^{l}\sqrt{\frac{l+l'+1}{2}}\sqrt{n!n'!\Gamma(n+l+3/2)\Gamma(n'+l'+3/2)} \\ 
&\times\sum_{m,m'=0}^{n,n'} \left\{
     \begin{array}{lr}
       \frac{(-1)^{m+m'}\left[2m\Gamma\left(m+m'+1+\frac{l+l'}{2}\right)-\Gamma\left(m+m'+1+\frac{l+l'}{2}\right)\right]}{m!m'!(n-m)!(n'-m')!\Gamma(m+l+3/2)\Gamma(m'+l'+3/2)} & \text{if}\: l'=l-1 \\
        \frac{(-1)^{m+m'+1}\left[(2m+2l+1)\Gamma\left(m+m'+1+\frac{l+l'}{2}\right)-\Gamma\left(m+m'+1+\frac{l+l'}{2}\right)\right]}{m!m'!(n-m)!(n'-m')!\Gamma(m+l+3/2)\Gamma(m'+l'+3/2)} & \text{if}\: l'=l+1 \\
       0 & \text{otherwise}
     \end{array}
   \right.
   \end{split}
\end{align}
If $l=1$ and $l'=0$ or vice versa, reduced matrix elements between one modified wave function of the form~\eqref{eq:BuschWF} and one pure harmonic oscillator state are needed. These are given by
\begin{align}
\braket{n l=0 || q || n' l'=1}=-i A(n) \sqrt{\frac{\Gamma(n'+5/2)}{2\pi^3 n'!}}\frac{2n-2n'-1}{2(n'-n)(1+n'-n)}
\end{align}
and its Hermitian conjugate.

Our choice of basis makes the relative matrix elements~\eqref{eq:WeylRel} simple at the cost of complicating the center-of-mass term. We take the approach of expanding the states~\eqref{eq:basisStates} in the alternate coupling scheme,
\begin{equation}\label{eq:basisStates2}
\ket{n(ls)j;NL;(jL)J}=(-1)^{l+s+L+J}\sqrt{2j+1}\sum_{\mathcal{J}}\sqrt{2\mathcal{J}+1}\sixj{l}{s}{j}{L}{J}{\mathcal{J}}\ket{nl;N(Ls)\mathcal{J};(l\mathcal{J})J}.
\end{equation}
Using this notation, the matrix elements can be written
\begin{equation}\begin{split}
\bra{n'(l's')j';N'L';(j'L')J'}&\vec{Q}\cdot\vec{\Sigma} \ket{n(ls)j;NL;(jL)J} = \delta_{n,n'}\delta_{l,l'}\delta_{J,J'}\delta_{s,1}\delta_{s1,1}   6 (-1)^{L} \\
&\hspace{-1cm}\times\braket{N'L'|| \vec{Q} || NL} \sum_{\mathcal{J}}(-1)^\mathcal{J} (2\mathcal{J}+1)\sixj{l}{1}{j'}{L'}{J}{\mathcal{J}}\sixj{l}{1}{j}{L}{J}{\mathcal{J}}\sixj{\mathcal{J}}{1}{L'}{1}{L}{1}.
\end{split}
\end{equation}
Again, the reduced matrix element of the center-of-mass momentum changes the parity by connecting states with $\Delta L=\pm1$. Matrix elements are nonzero only for $\Delta s=0$ because the antisymmetry of the spatial wave function depends only on $l$, which does not change. We also note that the c.m. term does not affect states with singlet spin wave functions ($s=0$).

\begin{figure}
\includegraphics{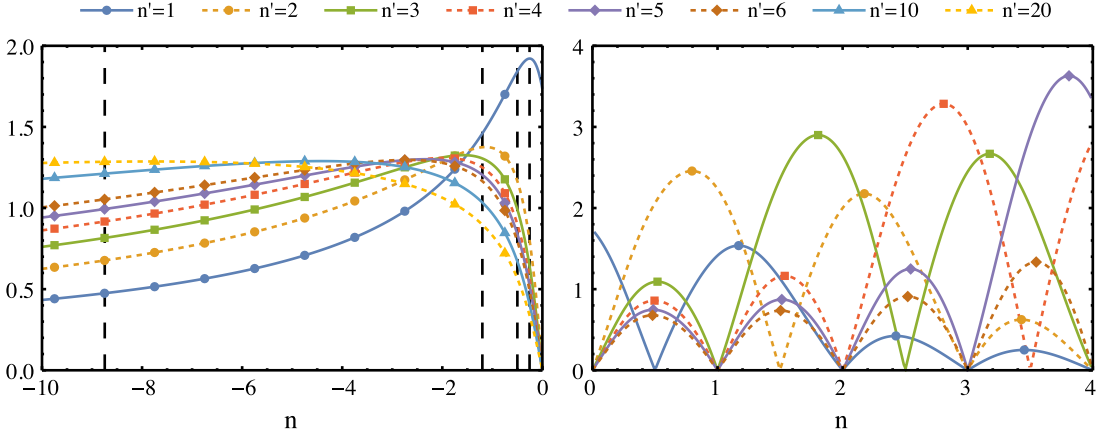}
\caption{\label{fig:matrixElts}(Color online)  Absolute value of the matrix elements $|\bra{n'(11)0;00;(00)0}\vec{\sigma}\cdot\vec{q}\ket{n(00)0;00;(00)0}|$ between the ground state and $l=1$ excited states. The horizontal axis is the principal quantum number of the ground state obtained by solving~\eqref{eq:eigenvalueEqn}. From left to right, the vertical lines on the negative axis indicate the values obtained for $\tilde{a}=1/4$, $\tilde{a}=1$, $\tilde{a}=\pm\infty$, and $\tilde{a}=-1$, respectively.} 
\end{figure}

Using these matrix elements, we calculated the spectrum of the two interacting particles with Weyl spin-orbit coupling. Our calculations are performed by numerically diagonalizing in a truncated basis of the harmonic oscillator states~\eqref{eq:basisStates}, where a cutoff $2N+L+2n+l+3\leq E_{\text{max}}$ is set high enough that the eigenvalues of the matrix have converged to the desired accuracy.  

\begin{figure}
\includegraphics{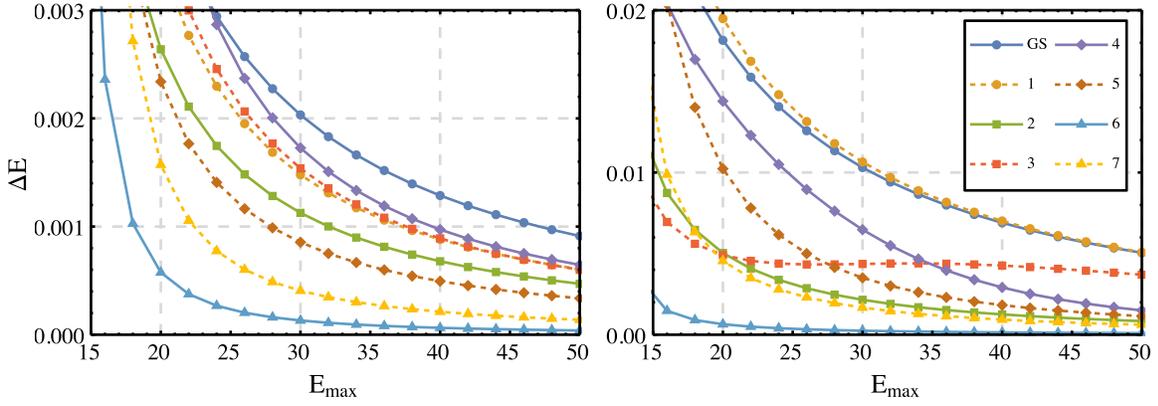}
\caption{\label{fig:WeylConvergence}(Color online)  A convergence plot giving the change in energy eigenvalue, $\Delta E$, for the lowest eight energy levels when a shell is added as a function of $E_{\text{max}}$.  The left figure shows convergence for $\tilde{a}=-1$ and $\tilde{\alpha}_W=0.5$. In the right panel  we show $\tilde{a}=1$ and $\tilde{\alpha}_W=0.5$, demonstrating that convergence of the states with large negative $n$ is poor.} 
\end{figure}

This approach converges well only when the ground-state energy is not too low. In particular, for $a$ positive but very small the principal quantum number of the ground state is increasing from negative infinity as seen in Fig.~\ref{fig:BuschSpectrum}. From Fig.~\ref{fig:matrixElts}, we can see that as $n$ becomes more negative, the principal quantum number of the dominant matrix element is also increasing. Because convergence of any energy level requires a cutoff much larger than the energy of the most strongly coupled states, a sufficiently high $E_{\text{max}}$ to ensure an accurate ground-state energy becomes infeasible for small positive $a$. For excited states, $n$ is always positive and matrix elements with similar $n$ always dominate. The strength of the matrix elements follows a similar qualitative behavior for the spin-orbit couplings treated in the following sections where the same issues recur. 

\begin{figure}
\includegraphics{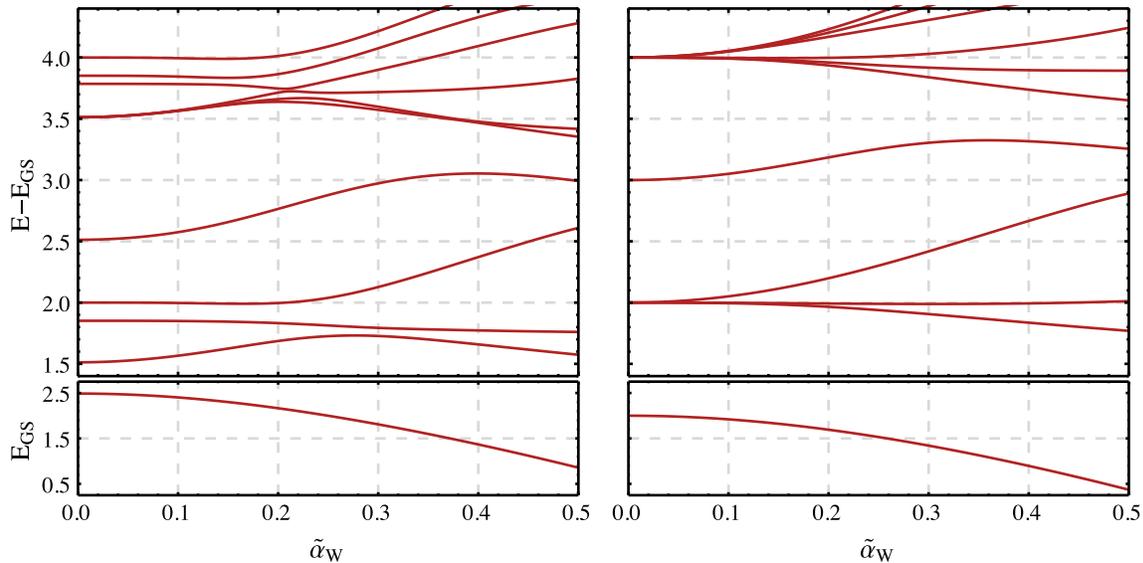}
\caption{\label{fig:WeylSpectrum}(Color online)  Spectrum of states with total angular momentum $J=0$ for the dimensionless Hamiltonian~\eqref{eq:WeylHamiltonian}. The bottom left figure shows the ground-state energy for $\tilde{a}=-1$ as a function of $\tilde{\alpha}_W$; above are the first few excitation energies. The right figure shows the results in the unitary limit of the two-body interaction, $|\tilde{a}|\rightarrow\infty$. The spectrum is symmetric about $\tilde{\alpha}_W=0$.} 
\end{figure}

As a result, convergence of the ground state is actually slower than that for nearby excited states. Furthermore, our approach gives the fastest convergence when $a$ is not small and positive. We compare the rate of convergence of the $\tilde{a}=-1$ and $\tilde{a}=1$ spectra in Fig.~\ref{fig:WeylConvergence} to demonstrate the dependence of convergence on the matrix truncation. The actual energy spectrum is shown in Fig.~\ref{fig:WeylSpectrum}. 

\begin{figure}
\includegraphics{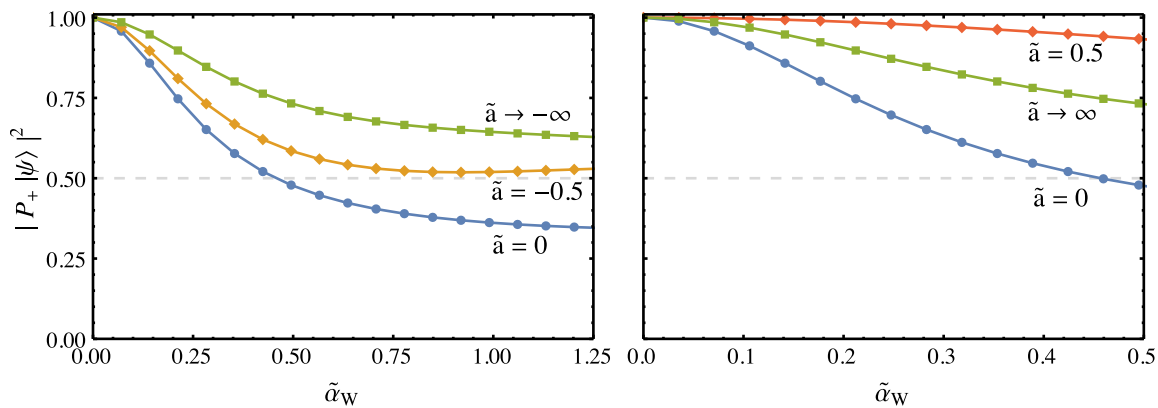}
\caption{\label{fig:Projections}(Color online) For different values of the two-body coupling strength $\tilde{a}$, we show the magnitude of the ground state projected onto even parity basis states as a function of the SOC strength. This is given by $\big|  P_+\ket{\psi_{\text{GS}}}\big|^2=\big| (1- P_-)\ket{\psi_{\text{GS}}}\big|^2$, where $P_+$ ($P_-$) is the projection operator onto the positive\nobreak- (negative\nobreak-) parity basis states. The left figure shows negative $\tilde{a}$, while the right shows positive $\tilde{a}$. Note that the limits $\tilde{a}\rightarrow\pm\infty$ are physically identical.}
\end{figure}

One consequence of parity violation in this system is that the eigenstates are mixtures of the even- and odd-parity basis states described by Eq.~\eqref{eq:basisStates}. In Fig.~\ref{fig:Projections} we visualize how these subspaces are mixed in the ground state as the SOC strength increases. For the noninteracting system, $\tilde{a}=0$, more than half of the ground state projects onto negative-parity states even at fairly small values of $\tilde{\alpha}_W$. However, we see that the short-range interaction reduces this effect. With negative $\tilde{a}$, the mixing of the negative-parity states is suppressed as the strength of the two-body interaction increases. When $\tilde{a}$ is positive the effect is more striking. Mixing with negative-parity states is most strongly suppressed for small positive values of $\tilde{a}$, while the projection onto these states increases for larger positive values. The admixture is qualitatively the same when considering other forms of SOC as described in the following sections.

\section{\label{sec:Rashba}The Pure Rashba Coupling}

In order to find the matrix elements of the pure Rashba coupling given in~\eqref{eq:Rashba}, we first note that it can be written as a spherical tensor
\begin{equation}
V_{R}=i\sqrt{2}\:\alpha_R \left[ k \otimes \sigma \right]_{10}.
\end{equation}
We therefore have the two-body Hamiltonian
\begin{equation}\label{eq:RashbaHamiltonian}
\frac{1}{\hbar\omega}H=\left(h_{0,\text{rel}}+i \tilde{\alpha}_R  \left[ \vec{q} \otimes \vec{\sigma} \right]_{10} + \sqrt{2}\pi \tilde{a}(\Lambda) \delta^{(3)}(r)\right)+\left(h_{0,\text{c.m.}}+i \tilde{\alpha}_R [ \vec{Q}\otimes \vec{\Sigma} ]_{10} \right).
\end{equation}

Because the spin-orbit coupling is now a $k=1$ tensor rather than a scalar operator, the total angular momentum $J$ is no longer conserved. Additionally, the matrix elements now depend on the quantum number $J_z$ (which is conserved). For the relative-coordinate part of the SOC, some algebra gives
\begin{equation}\begin{split}
&\bra{n'(l's')j';N'L';(j'L')J'J'_z}  [ \vec{q} \otimes \vec{\sigma} ]_{10}  \ket{n(ls)j;NL;(jL)JJ_z} =6 i (-1)^{J+J'-J'_z+j'+L+1}\delta_{N,N'}\delta_{L,L'}\delta_{J_z,J'_z} \\
 &\times\sqrt{(2J+1)(2J'+1)(2j+1)(2j'+1)} \threej{J'}{1}{J}{-J_z}{0}{J_z} \sixj{j'}{J'}{L}{J}{j}{1}
 \renewcommand{\arraystretch}{0.9}
 \ninej{\hphantom{l}l'\hphantom{l}}{\hphantom{l}l\hphantom{l}}{\hphantom{l}1\hphantom{l}}{s'}{s}{1}{j'}{j}{1} (s'-s) \braket{n'l' || q || n l}.
\end{split}
\end{equation}
For the center-of-mass part of the Hamiltonian we again expand the basis states in the alternate coupling scheme~\eqref{eq:basisStates2} to obtain the matrix elements
\begin{equation}\begin{split}
&\bra{n'(l's')j';N'L';(j'L')J'J'_z} [ \vec{Q} \otimes \vec{\Sigma} ]_{10}  \ket{n(ls)j;NL;(jL)JJ_z} = \delta_{n,n'}\delta_{l,l'}\delta_{J_z,J'_z}\delta_{s,1}\delta_{s',1} \\
 &\quad\times 6 i \sqrt{2}(-1)^{J+J'-J'_z+l} \sqrt{(2J+1)(2J'+1)(2j+1)(2j'+1)} \threej{J'}{1}{J}{-J_z}{0}{J_z}  \braket{N' L' || Q || N L} \\ 
 &\quad\times\sum_{\mathcal{J},\mathcal{J}'} (-1)^\mathcal{J}(2\mathcal{J}+1)(2\mathcal{J}'+1)\sixj{l}{1}{j'}{L'}{J'}{\mathcal{J}'}\sixj{l}{1}{j}{L}{J}{\mathcal{J}}\sixj{\mathcal{J}'}{J'}{l}{J}{\mathcal{J}}{1}
 \renewcommand{\arraystretch}{0.9}
 \ninej{\hphantom{l}L'\hphantom{l}}{\hphantom{l}L\hphantom{l}}{\hphantom{l}1\hphantom{l}}{1}{1}{1}{\mathcal{J}'}{\mathcal{J}}{1} .
\end{split}
\end{equation}

\begin{figure}
\includegraphics{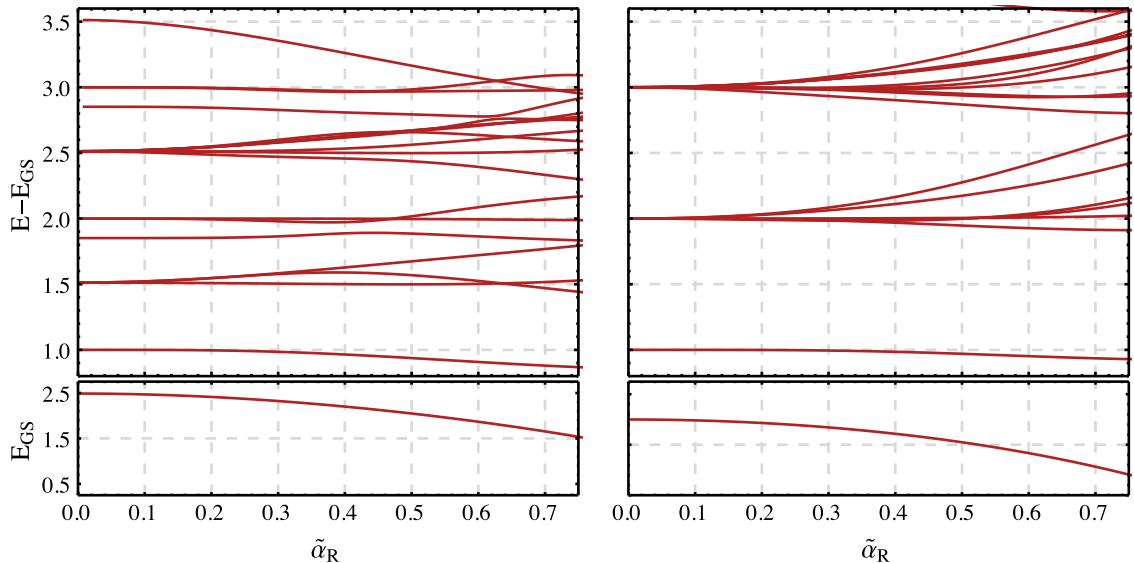}
\caption{\label{fig:RashbaSpectrum}(Color online)  Spectrum of states with total angular momentum quantum number $J_z=0$ for the Hamiltonian~\eqref{eq:RashbaHamiltonian}. The left figure shows the energies with negative scattering length $\tilde{a}=-1$. The right figure shows the results in the unitary limit $|\tilde{a}|\rightarrow\infty$. The spectrum is symmetric about $\tilde{\alpha}_R=0$.} 
\end{figure}

Our results for the Rashba SOC are shown in Fig.~\ref{fig:RashbaSpectrum}. Because the Rashba spin-orbit coupling is a vector operator, states of all possible $J$ must be included in any calculation and the size of the basis scales much more quickly with $E_{\text{max}}$. These spectra were computed with an $E_{\text{max}}$ of $24\hbar\omega$, for which there are approximately $36\,000$ basis states. All displayed eigenvalues of the Hamiltonian shift by less than $10^{-2}\hbar\omega$ if an additional shell of states is included.

\begin{figure}
\includegraphics{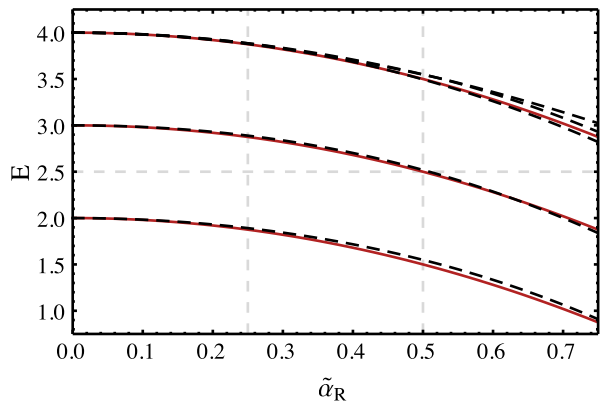}
\caption{\label{fig:ComparisonSpectrum}(Color online)  Comparison of selected spectral lines (dashed black) with the perturbative predictions from \cite{PhysRevA.89.033606} (solid red) when $\tilde{a}=\infty$. }
\end{figure}

\begin{figure}
\includegraphics{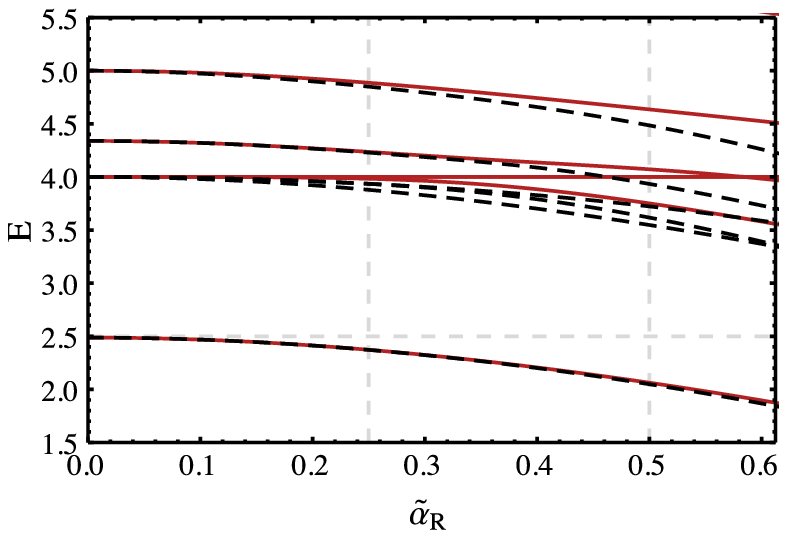}
\caption{\label{fig:ComparisonSpectrum2}(Color online)  A comparison of the energy levels with (dashed black) and without (solid red) the inclusion of excitations in the c.m. coordinate for $\tilde{a}=-1$. The approximation of ignoring c.m. excitations provides very accurate results for the ground state, but not for excited states.} 
\end{figure}

This interaction was also studied perturbatively for small $\alpha_R$ in \cite{PhysRevA.89.033606}, including the possibility of a spin-dependent two-body interaction, under the assumption that center-of-mass excitations are unimportant. For the specific case of identical fermions with spin-independent scattering length considered here, they found that the first correction to the energies occurs at order $\alpha_R^2$ and is independent of the scattering length $a$. We compare their perturbative predictions, which are derived from the non-degenerate theory, with our numerical results in Fig.~\ref{fig:ComparisonSpectrum}. 

By setting all matrix elements with $N,L>0$ in the bra or ket to zero, we also explored the approximation of ignoring center-of-mass excitations. Fig.~\ref{fig:ComparisonSpectrum2} shows that this is very accurate for the ground state, but less accurate for excited states. Suppression of the c.m. coordinate has a similar effect for the SOCs considered in Secs.~\ref{sec:Weyl} and~\ref{sec:R=D}. We also note that in the case of small positive $a$, the landscape of low-lying excited states is dominated by center-of-mass excitations. When $a\rightarrow0^+$ in the absence of spin-orbit coupling, there are an infinite number of states with nonzero c.m. quantum numbers whose energies lie between the ground state and the first relative-coordinate excitation.

\section{\label{sec:R=D} Equal-weight Rashba-Dresselhaus spin-orbit coupling}

Experiments have thus far realized only the effective Hamiltonian with equal strength Rashba and Dresselhaus couplings in the form~\eqref{eq:R=D}. Energy levels of the two-body system in the one-dimensional equivalent of this Hamiltonian with the additional magnetic field couplings present in experimental realizations have been calculated in \cite{guan2014energy}. Here we treat the problem in three dimensions.

This is also the most computationally difficult of the three cases. When decomposed into spherical tensors, the interaction~\eqref{eq:Dresselhaus} becomes
\begin{equation}
V_D=i\,\alpha_D \left( \left[ k \otimes \sigma \right]_{2,-2}- \left[ k \otimes \sigma \right]_{2,2}\right),
\end{equation}
and the two-particle Hamiltonian in the presence of equal strength Rashba and Dresselhaus SOC is given by~\eqref{eq:RashbaHamiltonian} with $\alpha_R\rightarrow \alpha_{R=D}$ plus the additional spin-orbit terms
\begin{equation}\label{eq:DresselhausHamiltonian}
\Delta H= \frac{i \tilde{\alpha}_{R=D}}{\sqrt{2}}\left(  \left[ \vec{q} \otimes \vec{\sigma} \right]_{2,-2} -  \left[ \vec{q} \otimes \vec{\sigma} \right]_{2,2} +[ \vec{Q} \otimes \vec{\Sigma} ]_{2,-2} -  [ \vec{Q} \otimes \vec{\Sigma} ]_{2,2} \right).
\end{equation} 
Yet again the number of basis states with nonzero matrix elements has increased; no angular momentum quantum numbers are conserved. The only remaining selection rule will be that the interaction does not change the total magnetic quantum number $J_z$ between even and odd. 

Using the same approach as in the previous sections, the matrix elements of the relative Dresselhaus term are
\begin{equation}\begin{split}
&\bra{n'(l's')j';N'L';(j'L')J'J'_z} \frac{i \tilde{\alpha}_{R=D}}{\sqrt{2}}\left(  \left[ \vec{q} \otimes \vec{\sigma} \right]_{2,-2} -  \left[ \vec{q} \otimes \vec{\sigma} \right]_{2,2} \right)  \ket{n(ls)j;NL;(jL)JJ_z}  \\
 &\quad\hphantom{\times}= i \sqrt{30}(-1)^{J+J'-J'_z+j'+L}\delta_{N,N'}\delta_{L,L'} \sqrt{(2J+1)(2J'+1)(2j+1)(2j'+1)}  \braket{n'l' || q || n l}\\
 &\hspace{1cm}\quad \times(s'-s) \left[\threej{J'}{2}{J}{-J'_z}{-2}{J_z}-\threej{J'}{2}{J}{-J'_z}{2}{J_z}\right] \sixj{j'}{J'}{L}{J}{j}{2}
 \renewcommand{\arraystretch}{0.9} \ninej{\hphantom{l}l'\hphantom{l}}{\hphantom{l}l\hphantom{l}}{\hphantom{l}1\hphantom{l}}{s'}{s}{1}{j'}{j}{2},
\end{split}
\end{equation}
while the center-of-mass part is 

\begin{equation}\begin{split}
&\bra{n'(l's')j';N'L';(j'L')J'J'_z}  \frac{i \tilde{\alpha}_{R=D}}{\sqrt{2}}\left(  \left[ \vec{Q} \otimes \vec{\Sigma} \right]_{2,-2} -  \left[ \vec{Q} \otimes \vec{\Sigma} \right]_{2,2} \right)  \ket{n(ls)j;NL;(jL)JJ_z}   \\
&\quad=2 i \sqrt{15}(-1)^{J+J'-J'_z+l+1}\delta_{n,n'}\delta_{l,l'}\delta_{s,1}\delta_{s',1}  \\
 &\quad\hphantom{=}\times \sqrt{(2J+1)(2J'+1)(2j+1)(2j'+1)} \left[\threej{J'}{2}{J}{-J'_z}{-2}{J_z}-\threej{J'}{2}{J}{-J'_z}{2}{J_z}\right] \braket{N' L' || Q || N L} \\ 
 &\quad\hphantom{=}\times\sum_{\mathcal{J},\mathcal{J}'} (-1)^\mathcal{J}(2\mathcal{J}+1)(2\mathcal{J}'+1)\sixj{l}{1}{j'}{L'}{J'}{\mathcal{J}'}\sixj{l}{1}{j}{L}{J}{\mathcal{J}}\sixj{\mathcal{J}'}{J'}{l}{J}{\mathcal{J}}{2}
 \renewcommand{\arraystretch}{0.9}
 \ninej{\hphantom{l}L'\hphantom{l}}{\hphantom{l}L\hphantom{l}}{\hphantom{l}1\hphantom{l}}{1}{1}{1}{\mathcal{J}'}{\mathcal{J}}{2} .
\end{split}
\end{equation}

\begin{figure}
\includegraphics{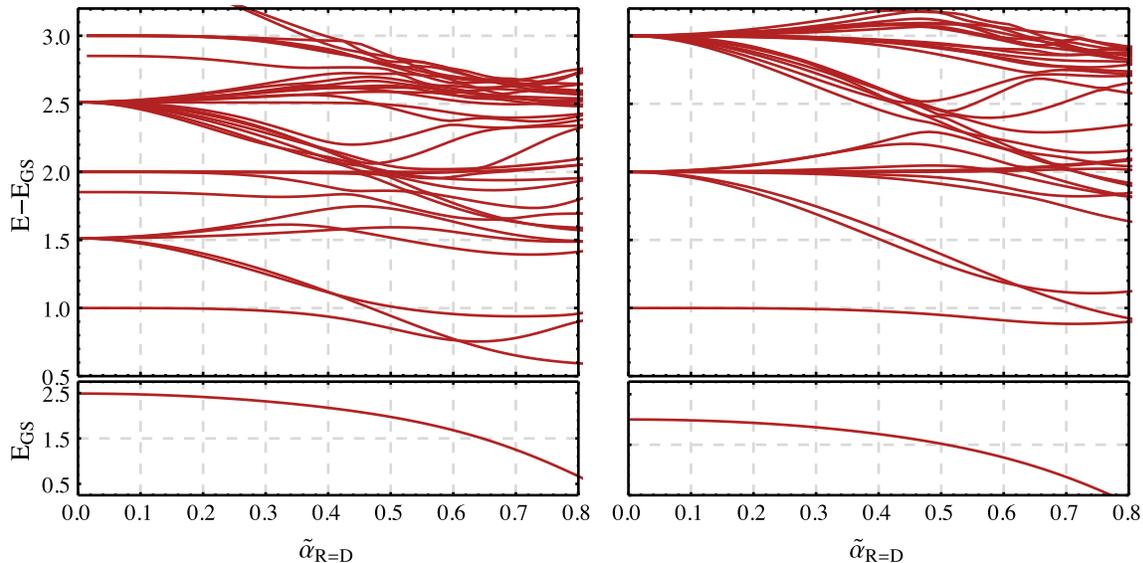}
\caption{\label{fig:R=DExcitationSpectrum} (Color online) 
Spectrum of states with even total angular momentum magnetic quantum number $J_z=0,2,\dots$ for the equal-weight Rashba-Dresselhaus SOC~\eqref{eq:R=D}. The left figure shows the energies with negative scattering length $\tilde{a}=-1$. The right figure shows the results in the unitary limit $|\tilde{a}|\rightarrow\infty$. The spectrum is symmetric about $\tilde{\alpha}_{R=D}=0$.} 
\end{figure}

The richly structured excitation spectrum of low-lying states is shown in Fig.~\ref{fig:R=DExcitationSpectrum} for a cutoff of $E_{\text{max}}=17$. All displayed energies shift by less than .$02\hbar\omega$ when the final shell is added, giving a slightly faster convergence than in the pure Rashba case.

\section{Conclusions}

In this paper we have nonperturbatively calculated the spectrum of interacting two-particle systems with realistic spin-orbit couplings when the trapping potential cannot be ignored.  Matrix elements of a short-range pseudopotential and three types of spin-orbit coupling were determined analytically in a basis of the total angular momentum eigenstates of the interacting two-body problem without SOC. With the analytic matrix elements, exact diagonalization of the Hamiltonian within a finite basis was possible.

Our energy calculations were performed in a basis truncated in a consistent way by including all states below an energy cutoff. The resulting spectra show good convergence except in the case where the two-body interaction generates a small positive scattering length. In this regime coupling of the ground state to higher relative-coordinate excited states dominates and convergence in the cutoff parameter $E_{\text{max}}$ was numerically intractable. We are currently investigating alternative methods to deal with this issue. In the limit of weak SOC we have compared our results to the perturbative calculations of \cite{PhysRevA.89.033606} and found good agreement. We also observed that although the ground state does not couple strongly to center-of-mass excitations, their inclusion is crucial for the excited state spectrum.  The relatively weak center-of-mass coupling of the ground state, however, suggests that cold atoms with SOC can be used as a surrogate system to probe properties of two-body spin-orbit couplings, e.g., the parity-violating weak interaction in nuclear systems. 

We provided plots of a variety of spectra calculated with Weyl, Rashba, and equal weight Rashba-Dresselhaus couplings.  Although in this paper we show spectra only within certain subspaces of conserved angular momentum quantum numbers, the approach presented is fully capable of generating results for all possible states. Larger SO-coupling constants are also accessible with larger basis sizes. The general method can easily be adapted to calculate energies for bosonic systems, or to new forms of SOC such as the recently proposed spin-orbital angular momentum coupling \cite{2014arXiv1411.1737S}.  

Using the eigenvectors of the truncated basis Hamiltonian, we also explored the effect of parity violation on the system. In particular we show how the SOC induces mixing of the positive- and negative-parity subspaces for the ground state. Without a two-body interaction, the ground state preferentially projects onto negative parity basis states even for modest SOC strength. The short-range interaction was seen to suppress this mixing, especially when the scattering length is positive.

A natural extension of this work is to consider three particles within a trap.  Because of the complex spectrum that is associated with three-body physics at the unitary limit (e.g., Efimov states, limit cycles, etc.), the spectrum under the influence of an external SOC is expected to be quite rich.  Couplings between the center-of-mass and relative motion due to the SOC present a potential challenge to traditional few-body techniques, such as the Faddeev equations, which work only within the relative coordinates.  However, in our two-body calculations we found that the coupling of the ground state to the c.m. motion is weak. If this is also true in the three-body case, then to a good approximation we can ignore the c.m. motion and utilize existing few-body techniques with little or no modification.  

\section*{Acknowledgements}
We thank Paulo Bedaque, Jordy de Vries and Timo L{\"a}hde for their input and discussions related to this work. This paper is based in part on work supported by the U.S. Department of Energy, Office of Science, Office of Nuclear Physics, under Award No. DE-SC00046548. 

\bibliography{SOCoupling}

\appendix*
\section{Derivation of the normalization factor for Busch wave functions}
In the original paper by Busch \textit{et al.} \cite{Busch}, the normalization factor of the wave functions is not given. The closed form expression for this normalization does not seem to be widely known. It was originally presented in \cite{PhysRevA.85.053614} without derivation, which we provide here. To find the norm of the wave function~\eqref{eq:BuschWF}, one must integrate (using a change of variables to $z=r^2$)
\begin{equation}\label{eq:NormIntegral}
A^{-2}=\frac{\Gamma(-n)^2}{8\pi^3}  \int_0^\infty \frac{1}{z}\left[U(-n,3/2,z)e^{-z/2} z^{3/4} \right]^2  dz.
\end{equation}
The term in brackets is equal to a Whittaker function \cite{DLMF} and so this can be rewritten,
\begin{equation}
A^{-2}=\frac{\Gamma(-n)^2}{8\pi^3}  \int_0^\infty \frac{1}{z}\left[W_{n+3/4,1/4}(z) \right]^2  dz.
\end{equation}
This integral can be found in \cite{GradshteynRyzhik}
\begin{equation}
\int_0^\infty \frac{1}{z}\left[W_{\kappa,\mu}(z) \right]^2  dz=\frac{\pi}{\sin (2\pi \mu)}\frac{\psi_0(\frac{1}{2}+\mu-\kappa)-\psi_0(\frac{1}{2}+\mu-\kappa)}{\Gamma(\frac{1}{2}+\mu-\kappa)\Gamma(\frac{1}{2}-\mu-\kappa)}.
\end{equation}
Applying this to~\eqref{eq:NormIntegral} with $\kappa=n+3/4$ and $\mu=1/4$ gives the desired result,
\begin{equation}
A^{-2}=\frac{1}{8\pi^3}  \frac{\Gamma (-n)}{\Gamma(-n-1/2)}\left[\psi_0(-n)-\psi_0(-n-1/2)\right].
\end{equation}

\end{document}